\documentclass[twocolumn,showpacs,preprintnumbers,amsmath,amssymb,prb,superscriptaddress]{revtex4-1}
\usepackage{epsfig}
\usepackage{graphicx}
\usepackage{dcolumn}
\usepackage{bm}
\usepackage{color} 

\begin{document}

\title{
Optical and photoelectrical studies on anisotropic metal-insulator transition of RuAs
}
\author{Yuki Nakajima}
\affiliation{Department of Physics, Graduate School of Science, Osaka University, Toyonaka 560-0043, Japan}
\author{Zenjiro Mita}
\affiliation{Graduate School of Frontier Biosciences, Osaka University, Suita 565-0871, Japan}
\author{Hiroshi Watanabe}
\affiliation{Graduate School of Frontier Biosciences, Osaka University, Suita 565-0871, Japan}
\affiliation{Department of Physics, Graduate School of Science, Osaka University, Toyonaka 560-0043, Japan}
\author{Yoshiyuki Ohtsubo}
\affiliation{Graduate School of Frontier Biosciences, Osaka University, Suita 565-0871, Japan}
\affiliation{Department of Physics, Graduate School of Science, Osaka University, Toyonaka 560-0043, Japan}
\author{Takahiro Ito}
\affiliation{Nagoya University Synchrotron Radiation Research Center (NUSR), Nagoya University, Nagoya 464-8603, Japan}
\affiliation{Graduate School of Engineering, Nagoya University, Nagoya 464-8603, Japan}
\author{Hisashi Kotegawa}
\affiliation{Department of Physics, Graduate School of Science, Kobe University, Kobe 657-8501, Japan}
\author{Hitoshi Sugawara}
\affiliation{Department of Physics, Graduate School of Science, Kobe University, Kobe 657-8501, Japan}
\author{Hideki Tou}
\affiliation{Department of Physics, Graduate School of Science, Kobe University, Kobe 657-8501, Japan}
\author{Shin-ichi Kimura}
\email{kimura@fbs.osaka-u.ac.jp}
\affiliation{Graduate School of Frontier Biosciences, Osaka University, Suita 565-0871, Japan}
\affiliation{Department of Physics, Graduate School of Science, Osaka University, Toyonaka 560-0043, Japan}
\date{\today}
\begin{abstract}
The anisotropic changes in the electronic structure of a metal-to-insulator transition (MIT) material, RuAs, with two-step phase transition are reported by using polarized optical conductivity [$\sigma(\omega)$] spectra, angle-integrated photoelectron (PE) spectra, and band calculations based on local density approximation (LDA).
Both the PE and $\sigma(\omega)$ spectra not only in the high-temperature (HT) phase but also in the low-temperature (LT) phase 
as well as the energy gap formation owing to the MIT were almost consistent with those derived from the LDA band calculations, 
so the fundamental electronic structure in the HT and LT phases can be explained without electron correlations.
However, the electronic structure in the middle phase between the HT and LT phases has not been clarified.
The polarized $\sigma(\omega)$ spectra revealed not only the anisotropic energy gap formation but also the anisotropic gap-opening temperature, i.e., 
the energy gap along the $c$ axis in the HT phase starts to open near the higher transition temperature,
but that along the $b$ axis opens below the lower transition temperature.
The finding suggests that the two-step MIT originates from the anisotropic energy gap formation.
\end{abstract}

%
\maketitle
%
\section{Introduction}
After the discovery of iron pnictides (Fe-$X$; $X=$ P, As, Sb) superconductors~\cite{Kamihara2008}, physical properties of related materials have been widely investigated to obtain hints to solve the mechanism of the superconductivity.
The substitution of Fe to ruthenium (Ru), which is one row below Fe in the periodic table, has been performed, and Ru$_{1-x}$Rh$_x$As, Ru$_{1-x}$Rh$_x$P, and RuSb show superconductivity below several kelvin (K).~\cite{Hirai2012}
On the other hand, pristine RuAs and RuP that have an orthorhombic MnP-type structure with the $Pnma$ space group (No.~62)~\cite{Rundqvist1962,Saparov2012} at room temperature show a metal-to-insulator transition (MIT) with decreasing temperature accompanied with structural change.
The MIT is a first order transition from the paramagnetic phase at high temperature to the nonmagnetic phase at low temperature~\cite{Kuwata2018}, suggesting the formation of charge-density waves (CDWs).

In the series of ruthenium mono-pnictides, optical conductivity [$\sigma(\omega)$] spectra~\cite{Chen2015}, angle-integrated photoelectron (PE) spectra~\cite{Sato2012}, and NMR~\cite{Fan2015,Li2017} of RuP have been investigated to clarify the electronic structure~\cite{Chen2015,Sato2012}.
The electrical resistivity data of this material suggest that the metallic phase at room temperature changes to the insulating phase with spin gap below the temperature of $\sim250$~K~\cite{Li2017}.
The $\sigma(\omega)$ spectra and NMR suggest that the origin of the phase transition is the CDW formation, but no clear band nesting has been observed in PE spectra.
Sato {\it et al.} proposed another possibility of the successive Peierls-like transitions through the different orbitals~\cite{Sato2012}.
In addition, the direct current conductivity ($\sigma_{DC}$) of the CDW state is inconsistent in two groups:
Hirai {\it et al.} have reported that the CDW state was insulator with $\sigma_{DC}$ of $1~\Omega^{-1}$cm$^{-1}$ below 10~K~\cite{Hirai2012}, 
but Chen {\it et al.} have shown a metallic character with $\sigma_{DC}\sim0.1$~m$\Omega^{-1}$cm$^{-1}$ at 10~K by an electrical resistivity and $\sigma(\omega)$ data~\cite{Chen2015}.
In the PE spectra with 10-eV photons, which is considered to reflect bulk electronic structures mainly, the intensity at the Fermi level ($E_{\rm F}$) has not been changed so much with decreasing temperature across the phase transition temperature~\cite{Sato2012}.
Therefore, the origin of the electronic structure change at the phase transition temperature is still under debate.

Recently, single crystals of RuAs that show similar physical properties to RuP have been grown and used for the investigation of the phase transition~\cite{Kotegawa2018}.
The metallic character at room temperature changes to an insulating one detected by an electrical resistivity measurement through a two-step structural phase transition at about 250~K ($T_{MI1}$) and about 200~K ($T_{MI2}$).
The orthorhombic crystal structure with the space group of $Pnma$ at room temperature change to the formation of a superlattice of $3\times3\times3$ of the original unit cell of monochrinic structure with the space group of $P2_1/c$ (No.~14) below $T_{MI2}$.
The electronic structure as well as theoretical polarized $\sigma(\omega)$ spectrum at the high temperature (HT) phase above $T_{MI1}$ has been obtained by band calculations based on local density approximation (LDA)~\cite{Goto2015}.
Recently, the crystal structure in the low temperature (LT) phase below $T_{MI2}$ has been revealed by X-ray diffraction, and the LDA band structure calculations in the LT phase have been also performed~\cite{Kotegawa2018}.
The band calculations show that the density of states (DOS) at $E_{\rm F}$ decreases from that in the HT phase but does not disappears in the LT phase, which is inconsistent with the insulating electrical resistivity ($\sim1~\Omega$cm) at the lowest accessible temperature.

The two-step phase transition commonly appears in both RuAs and RuP, so it may be important for the physical properties of Ru mono-pnictides.
In comparison with the HT and LT phases, the middle-temperature (MT) phase at temperatures between $T_{MI1}$ and $T_{MI2}$ has not been elucidated.
By using X-ray diffraction, the crystal structure in the MT phase looks like incommensurate, and multi domains appear.
Therefore, since X-ray diffraction is not applicable to reveal the MT phase, 
it is necessary to use other observation methods, in particular, direct observation of changes in electronic structure.

In this paper, 
to clarify the change of the electronic structure of RuAs from the HT phase to the LT phase as well as the origin of the phase transition, 
we have measured the temperature dependence of polarized $\sigma(\omega)$ spectra along the $b$ and $c$ axes in the HT phase and the PE spectrum, 
and compared them with LDA band calculations.
In the HT phase, PE spectra of the wide valence band and polarized $\sigma(\omega)$ spectra in the interband transition region could be explained well by the LDA band calculations.
The PE intensity near $E_{\rm F}$ in the LT phase was lower than that in the HT phase, which can be also explained by the different DOS near $E_{\rm F}$ in the HT and LT phases.
Anisotropic energy gap structures were observed in the $\sigma(\omega)$ spectra in the LT phase, which can be also explained well by the LDA calculations.
The energy gap size and also the gap-opening temperature were anisotropic, which suggest an anisotropic CDW/charge-ordering formation. 

\section{Experimental}
%
\begin{figure}[t]
\begin{center}
\includegraphics[width=0.40\textwidth]{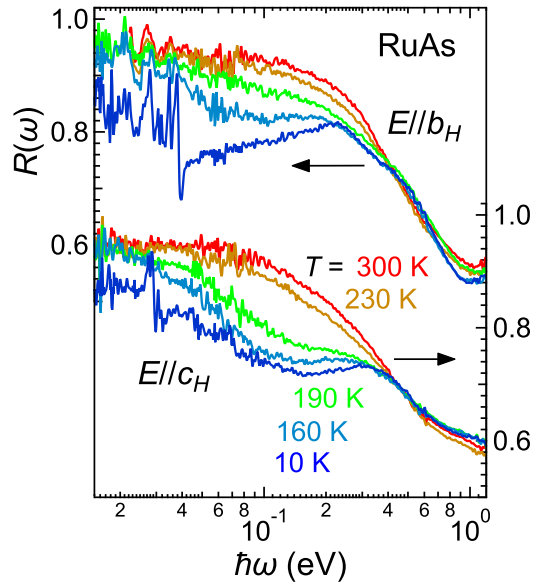}
\end{center}
\caption{
Temperature-dependent polarized reflectivity [$R(\omega)$] spectra of RuAs along the $b$ and $c$ axes in the HT phase (namely $b_H$ and $c_H$ axes, respectively) in the photon energy $\hbar\omega$ range of 0.015--1.2~eV.
Fine structures in the photon energy region below 0.04~eV originate from TO phonons (not discussed here).
}
\label{fig:reflectivity}
\end{figure}

Single-crystalline RuAs samples with the largest size of about $0.1\times0.5\times0.2$~mm$^3$ were synthesized by the Bi-flux method~\cite{Kotegawa2018}.
The longest crystal length of as-grown crystals was along the [011] direction, 
so as-grown surfaces along the $bc$ plane were measured for the optical reflectivity [$R(\omega)$] spectra.
Near-normal-incident polarized $R(\omega)$ spectra along the $b$ and $c$ axes were acquired in a wide photon-energy range of 15~meV -- 30~eV to ensure accurate Kramers-Kronig analysis (KKA)~\cite{Kimura2013}.
Infrared (IR) and terahertz (THz) measurements at the photon energy $\hbar\omega$ regions of 15~meV--1.5~eV have been performed using conventional reflectivity measurement setups to obtain absolute reflectivity with the accuracy of $\pm1-2$~\% with a feed-back positioning system in the temperature range of 10--300~K~\cite{Kimura2008}.
The accuracy of $R(\omega)$ is not as high as our normal $R(\omega)$ measurements~\cite{Kimura2017} because of the tiny sample size.
To obtain the absolute $R(\omega)$ values, the {\it in-situ} gold evaporation method was adopted.
IR and THz microscopes using synchrotron radiation at the beamline 6B of UVSOR-III, Institute for Molecular Science, Japan, have been also used~\cite{Kimura2007}.
The obtained temperature-dependent $R(\omega)$ spectra of RuAs along the $b$ and $c$ axes are plotted in Fig.~\ref{fig:reflectivity}.
In the photon energy range of 1.2--30~eV, the $R(\omega)$ spectrum was measured only at 300~K by using the synchrotron radiation setup at the beamline 7B of UVSOR-III, and connected to the spectra for $\hbar\omega \leq 1.5$~eV for KKA.
In order to obtain $\sigma(\omega)$ via KKA of $R(\omega)$, 
the spectra were extrapolated below 15~meV with a Hagen-Rubens function [$R(\omega)=1-\{2\omega/(\pi\sigma_{DC})\}^{1/2}$] for metallic cases at temperatures of $T \geq 210$~K and with a constant for insulating cases at temperatures of $T \leq 190$~K, 
and above 30~eV with a free-electron approximation $R(\omega) \propto \omega^{-4}$~\cite{DresselGruner2002}.
The extrapolation of the low-energy side was confirmed not to affect to the $\sigma(\omega)$ spectra at around 100~meV so much, which is the main part in this paper.

PE spectrum of the valence band at 200~K and temperature-dependent PE spectra near $E_{\rm F}$ have been measured using 100-eV photons from the synchrotron radiation beamline 7U of Aichi Synchrotron Radiation Center and 10-eV photons from the beamline 7U of UVSOR-III~\cite{Kimura2010}, respectively.
Samples have been cleaved perpendicular to the $[011]$ direction under ultra-high vacuum.
The $E_{\rm F}$ was calibrated by the Fermi edge of Au polycrystalline films electrically attached to the sample.

LDA band structure calculations have been performed by using the {\sc Wien2k} code including spin-orbit coupling~\cite{Blaha1990} to explain the experimentally obtained PE and $\sigma(\omega)$ spectra.
Lattice parameters of the HT and LT phases shown in Ref.~\cite{Kotegawa2018} were adopted and the obtained band structures were consistent with those in the paper.
Owing to the structural phase transition, the lattice space group of $Pnma$ in the HT phase above $250$~K changes to that of $P2_1/c$ in the LT phase below 200~K, 
where the directions of the primitive axes are changed, 
i.e., angle between the $a$ axis of the HT phase (namely $a_H$ axis, hereafter) and the $a$ ($c$) axis of the LT phase is 29.974$^\circ$ (49.183$^\circ$), 
and the $c$ axis of the HT phase (namely $c_H$ axis) is parallel to the $b$ axis of the LT phase.
The crystal axes in HT phase ($a_H$, $b_H$, and $c_H$) are parallel to $0.570a_L+0.430c_L$, $0.534a_L+0.403c_L$, and $b_L$ in the LT phase, respectively.
Here, we use the axes of the HT phase to discuss the anisotropic $\sigma(\omega)$ spectra.

\section{Results and Discussion}

\subsection{Valence band electronic structure}
\begin{figure}[t]
\begin{center}
\includegraphics[width=0.40\textwidth]{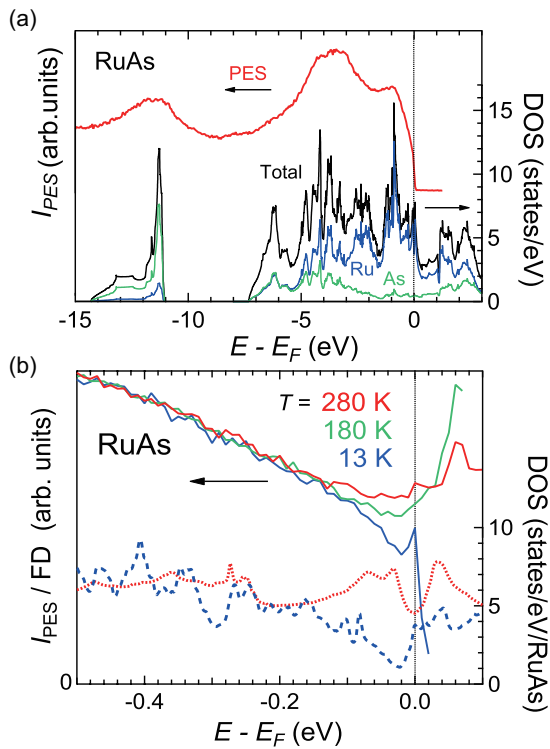}
\end{center}
\caption{
(a) Angle-integrated photoelectron (PE) spectra of the valence band ($I_{PES}$) measured at 200~K using 100-eV photons compared with the density of states (DOS) obtained from the LDA calculations.
(b) Temperature-dependent PE spectra near the Fermi level using 10-eV photons divided by the Fermi-Dirac distribution functions at the corresponding temperatures with instrumental energy resolution ($I_{PES}/FD$) normalized at $E-E_{\rm F}=-0.5$~eV compared with the DOS's at the high-temperature phase (dotted line) and the low-temperature phase (dashed line).
}
\label{fig:AIPES}
\end{figure}
Firstly, we compare the valence-band PE spectrum measured at the temperature of 300~K with the density of states (DOS) derived from band calculations of the HT phase shown in Fig.~\ref{fig:AIPES}(a).
In the PE spectrum, a cut-off at $E_{\rm F}$, peaks at $E-E_{\rm F}$ of -1, -3.5, and -11.5~eV, and a shoulder at $-6$~eV are visible.
These structures can be reproduced in the DOS, i.e., peaks from -7 to 0~eV mainly belong to the Ru ions, and the peak at -11.5~eV mainly originates from the As~$4s$ state.
The energies of these experimental peaks and other structure are similar to those of DOS, i.e., peak shifts owing to self-energy effects, which are usually observed in strongly correlated electron systems, do not appear, suggesting weak electron correlations~\cite{Hewson1993}.

\begin{figure}[t]
\begin{center}
\includegraphics[width=0.45\textwidth]{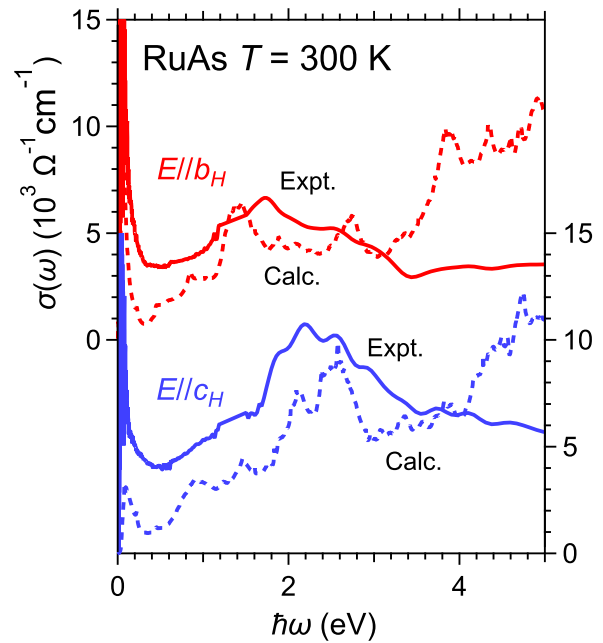}
\end{center}
\caption{
Anisotropic optical conductivity [$\sigma(\omega)$] spectra with $E \parallel c_H$ and $E \parallel b_H$ in the photon energy range below 5~eV (solid lines) compared with the calculated $\sigma(\omega)$ spectra (dashed lines).
}
\label{fig:OC_VIS}
\end{figure}
Polarized $\sigma(\omega)$ spectra near the visible region at 300~K are shown in Fig.~\ref{fig:OC_VIS}.
We can find that peaks appear at 1.7 and 2.1~eV in $E \parallel b_H$ and $E \parallel c_H$, respectively.
Calculated $\sigma(\omega)$ spectra in the HT phase shown in the figure have peaks at $\sim1.4$~eV in $E \parallel b_H$ and at $\sim2.5$~eV in 
$E \parallel c_H$, which seem to reproduce the experimental anisotropic spectra.
Therefore, high energy $\sigma(\omega)$ spectra shown in Fig.~\ref{fig:OC_VIS} also suggests that the electronic structure of RuAs can be explained by a simple LDA band calculations without any electron correlation. 

%
\subsection{Electronic structure change due to the phase transition}

Next, we show the change of the electronic structure from the HT phase to the LT phase (except for the MT phase).
Temperature-dependent PE spectra near $E_{\rm F}$ are shown in Fig.~\ref{fig:AIPES}(b).
The spectra are divided by the Fermi-Dirac distribution curves with corresponding temperatures convolved with instrumental energy resolution 
and normalized by the intensity at $E-E_{\rm F}=-0.5$~eV to show the temperature dependence of DOS.
At 280~K, the spectrum near $E_{\rm F}$ is almost flat suggesting metallic conductions,
but, with decreasing temperature, the intensity at $E_{\rm F}$ decreases.
This temperature dependence is consistent with the increase of the electrical resistivity at lower temperature, i.e., the carrier density decreases with decreasing temperature.
Although the electrical resistivity changes by several orders of magnitude from room temperature to the lowest accessible temperature, the PE intensity at $E_{\rm F}$ changes slightly.
It may originate from the surface sensitivity of PE spectra, i.e., the surface state of RuAs is metallic and it hardly changes even if the temperature is changed.
The low-temperature spectrum deviates from the high temperature spectrum in the energy range of $E-E_{\rm F} \leq -0.15$~eV.
The onset energy of $-0.15$~eV is almost the same as the energy 
at which the change of DOS from the HT phase to the LT phase as shown in the lower part of Fig.~\ref{fig:AIPES}(b).
Therefore, the temperature dependence near $E_{\rm F}$ in the PE spectrum is concluded to originate from MIT.

Temperature-dependent PE spectra of RuP, a related material to RuAs, suggests that
the energy gap appears below the energy $E-E_{\rm F}$ of about 50~meV, which is about three times smaller than the gap size observed in RuAs~\cite{Sato2012}.
In RuP, because of two phase transition temperatures of 330 and 270~K, the ratio of the gap size to the critical temperature becomes $2\Delta/k_BT_c \sim 3$, suggesting a Peierls-like transition~\cite{Gruner1994}.
However, in RuAs, the gap size is larger but the critical temperature is lower, then $2\Delta/k_BT_c\sim19$, which is not a reasonable value for a Peierls-like transition.
This value suggests that the energy gap originates not only from a Peierls-like transition but also from other ordering effect, such as a charge ordering.

\begin{figure}[t]
\begin{center}
\includegraphics[width=0.40\textwidth]{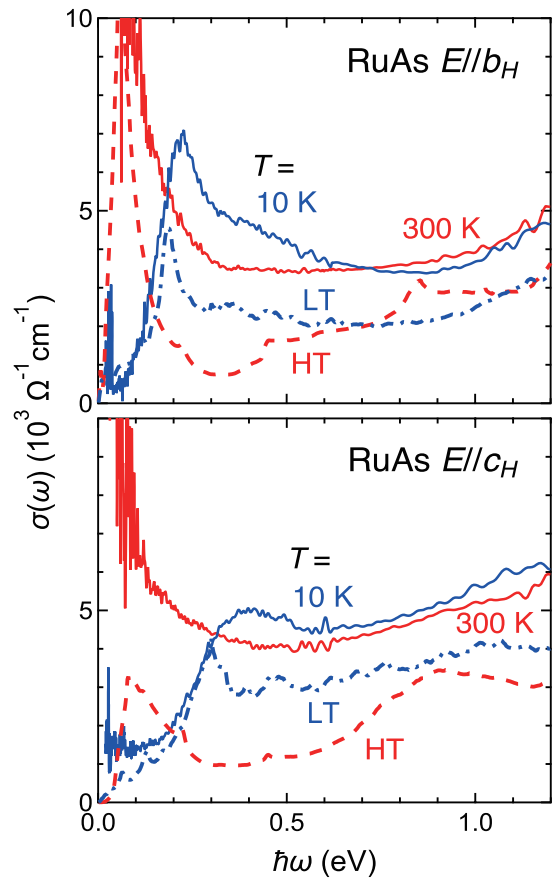}
\end{center}
\caption{
Optical conductivity [$\sigma(\omega)$] spectra (solid lines) at $T=300$~K (HT phase) and $10$~K (LT phase) compared with the corresponding interband spectra obtained from the LDA band calculations (dashed lines for the HT phase and dot-dashed lines for the LT phase) along the $b$ axis ($E \parallel b_H$) and $c$ axis ($E \parallel c_H$) of the HT phase.
}
\label{fig:CDWpeak}
\end{figure}
Polarized $\sigma(\omega)$ spectra in the IR region at 300~K (HT phase) and at 10~K (LT phase) are shown in Fig.~\ref{fig:CDWpeak}.
In both directions of $E \parallel b_H$ and $E \parallel c_H$, Drude-like $\sigma(\omega)$ spectra, 
in which the $\sigma(\omega)$ intensity increases with decreasing $\hbar\omega$, 
appear in the $\hbar\omega$ region below 0.3--0.4~eV at 300~K.
At 10~K, however, the Drude-like structures disappear and, instead, energy gaps with peaks appear at about 0.22~eV (0.4~eV) in $E \parallel b_H$ ($E \parallel c_H$), i.e., in the LT phase, energy gaps open along both crystal axes, but their sizes are different.

Calculated $\sigma(\omega)$ spectra obtained from the LDA calculations are also plotted by dashed and dot-dashed lines for the HT and LT phases, respectively.
The calculated spectra can reproduce well the spectral change from the HT phase to the LT phase.
In $E \parallel b_H$, a sharp peak in the calculation appears at $\hbar\omega\sim0.18$~eV, which corresponds to the experimental sharp peak at 0.22~eV, a broad hump in the $\hbar\omega$ region of $0.25$--$0.9$~eV and upturn with increasing $\hbar\omega$ above 0.9~eV in the calculation correspond to a tender slope and upturn, respectively, in the experiment at the same energy region.
At 300~K, the experimental spectral shape is also similar to the calculated one.
In addition, the spectral intensity at 10~K is larger than that at 300~K in the $\hbar\omega$ region of 0.2--0.7~eV, which is reproduced in the calculated spectra in 0.15--0.7~eV.
In $E \parallel c_H$, the experimental $\sigma(\omega)$ spectra can be also reproduced by the calculation, i.e., 
the experimental broad peak at 0.4~eV at 10~K corresponds to a calculated peak at 0.3~eV, 
and the calculated peak shape is broader than that in $E \parallel b_H$, which is the same relation in the experimental spectra.
The spectral intensity at 10~K is larger than that at 300~K in the $\hbar\omega$ region above 0.3~eV, which is also consistent with the calculated spectra although in the $\hbar\omega$ region above 0.2~eV.
These results suggest that the peaks at 0.2~eV in $E \parallel b_H$ and at 0.4~eV in $E \parallel c_H$ originate from the change of the superlattice formation of $3\times3\times3$ period in the LT phase.
Therefore, the difference of the experimental spectrum at 10~K from that at 300~K can be realized in the band calculations suggesting that the LDA band calculations are consistent with the obtained polarized $\sigma(\omega)$ spectra.
This result is also consistent with that derived from the PE spectra.

It should be noted that experimental $\sigma(\omega)$ spectra at 300~K have upturn structure Drude part originating from carriers, but in the calculation, interband transition components were only calculated without Drude components.
Nevertheless, the experimental spectral feature below 0.3~eV can be realized by the calculation.
This implies that interband transition components overlay on the Drude components in the HT phase.
The low-energy shift of the peak from the LT phase to the HT phase will be related to the MIT.
The peak shift is discussed below in detail.

\subsection{Temperature-dependent $\sigma(\omega)$ spectra}

\begin{figure}[t]
\begin{center}
\includegraphics[width=0.35\textwidth]{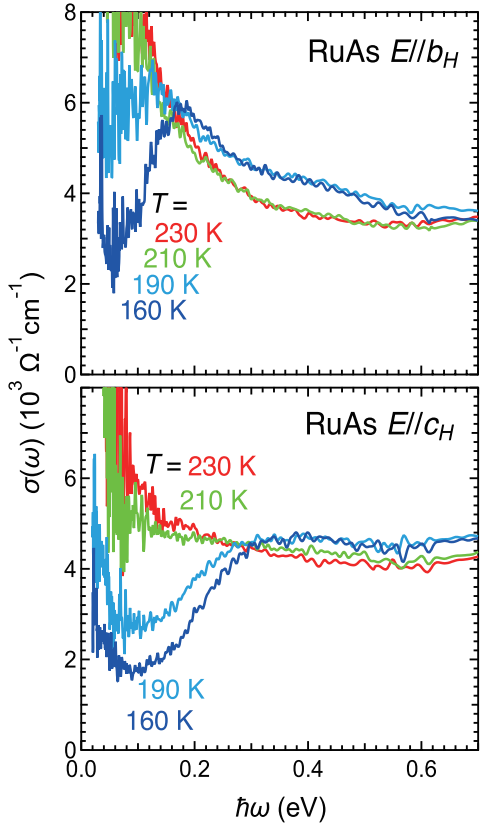}
\end{center}
\caption{
Temperature-dependent polarized $\sigma(\omega)$ spectra of RuAs in $E \parallel b_H$ and $E \parallel c_H$ in the photon energy range below 0.7~eV near the first-order phase transition temperature of $\sim200$~K ($T_{MI2}$).
}
\label{fig:TdepOC}
\end{figure}
To clarify the change of the electronic structure due to the phase transition, 
we have measured the detail temperature dependence of the $\sigma(\omega)$ spectrum. 
Figure \ref{fig:TdepOC} shows the temperature dependence of the $\sigma(\omega)$ spectra at around 200~K ($\sim T_{MI2}$) where large spectral change is expected. 
In $E \parallel b_H$, the spectrum changes little above 200~K, but below the temperature, the intensity at $\sim0.1$~eV decreases and that at $\hbar\omega \leq 0.16$~eV increases suggesting the appearance of an energy gap.
At 160~K, the energy gap structure becomes recognized clearly as shown in Fig.~\ref{fig:TdepOC}.
On the other hand, in $E \parallel c_H$, the spectrum at 210~K slightly changes from that at 230~K, where the intensity below (above) 0.22~eV decreases (increases).
The change becomes clearer with decreasing temperature, which indicates the energy gap opening.
These results imply that the spectral change in $E \parallel b_H$ starts at $T \leq 210$~K, but that in $E \parallel c_H$ gradually occurs between 210 and 230~K.

\begin{figure}[t]
\begin{center}
\includegraphics[width=0.45\textwidth]{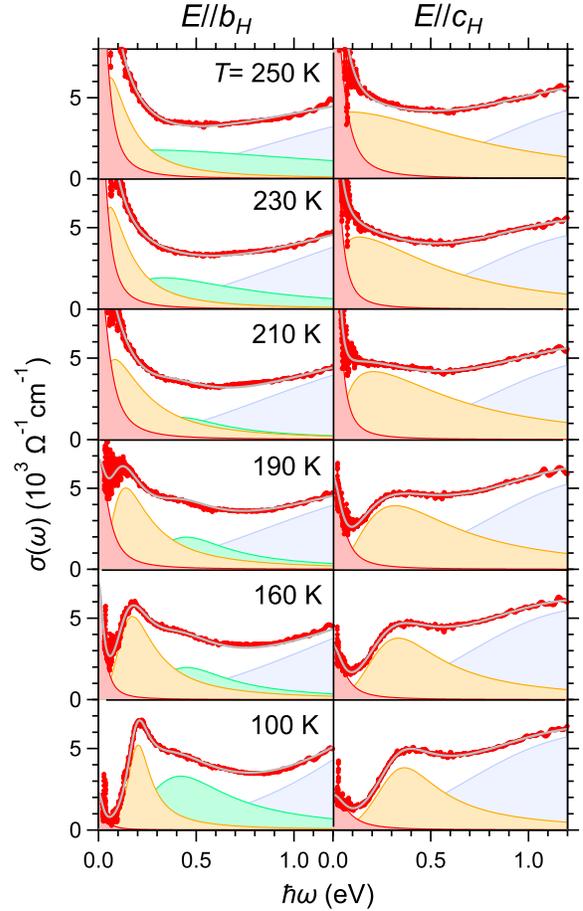}
\end{center}
\caption{
Temperature-dependent $\sigma(\omega)$ spectra of RuAs in $E \parallel b_H$ and $E \parallel c_H$ 
with the Drude and Lorentz fitting results.
One Drude and three (two) Lorentz functions are used for the fitting in $E \parallel b_H$ ($E \parallel c_H$).
The lowest Lorentz peak corresponds to the lowest excitation between the energy gap appearing in the LT phase. 
}
\label{fig:fitting}
\end{figure}
To clarify the above inference，the $\sigma(\omega)$ spectra have been fitted by using the combination of Drude and Lorentz functions as follows~\cite{Wooten1972}; 
\begin{equation}
\sigma(\omega)=\dfrac{N_D e^2\Gamma_D}{m[\omega^2+\Gamma_D^2]}
+\sum_{i}\frac{N^*_ie^2\tau_i\omega^2}{m[(\omega^2_i-\omega^2)^2\tau^2_i+\omega^2]} . \nonumber
\end{equation}
Here, $m$ and $e$ denote the rest mass and the charge of an electron, respectively,
$N_D$ and $\Gamma_D$ are the effective carrier number and damping constant of carriers, respectively, of the Drude component and $\omega_i$, $N^{*}_i$, and $\Gamma_i$ are the peak energy, the effective electron number, and the damping constant of excited electrons, respectively, of each Lorentz component.

The fitting results mainly in the MT and LT phases are shown in Fig.~\ref{fig:fitting}.
Because of spectral shapes, the combinations of one Drude and three Lorentz functions in $E \parallel b_H$ and of one Drude and two Lorentz functions in $E \parallel c_H$ for the fitting functions were adopted.
The Drude weights along both directions commonly decrease with decreasing temperature, which is consistent with the increase of the electrical resistivity as well as the temperature-dependent PE spectra.

Here, we focus on the lowest-energy Lorentz peaks at $\hbar\omega\sim0.2$~eV in $E \parallel b_H$ and at about 0.32~eV in $E \parallel c_H$ at 100~K.
With decreasing temperature, both the peaks gradually deviate from $\hbar\omega=0$~eV with decreasing temperature suggesting the energy-gap opening in the LT phase, 
which corresponds to the calculated $\sigma(\omega)$ spectra in Fig.~\ref{fig:CDWpeak}.
Therefore, the temperature dependence of the lowest Lorentz peak is considered to have information of the phase transition.

\begin{figure}[t]
\begin{center}
\includegraphics[width=0.45\textwidth]{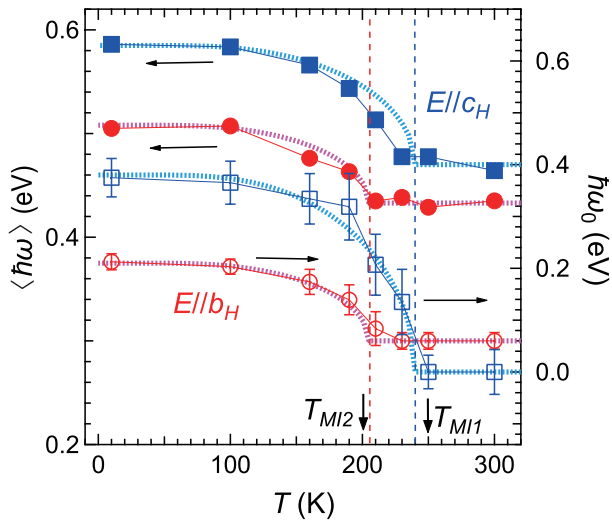}
\end{center}
\caption{
Temperature-dependent peak energy ($\hbar\omega_0$) and the center of gravity ($\langle\hbar\omega\rangle$) of $\sigma(\omega)$ spectra in $E \parallel b_H$ and $E \parallel c_H$.
Open circles (open squares) indicate the peak energy of the lowest-energy Lorentz peak in $E \parallel b_H$ ($E \parallel c_H$) shown in Fig.~\ref{fig:fitting} (right scale).
Solid circles (solid squares) are centers of gravity of $\sigma(\omega)$ spectra in $E \parallel b_H$ ($E \parallel c_H$) evaluated in the photon energy range of 0--1.2~eV (left scale).
Thick dotted lines are the temperature-dependent energy gaps predicted by BCS theory with the transition temperatures of $205$~K ($E \parallel b_H$) and $240$~K ($E \parallel c_H$).
Vertical dashed lines indicating the assumed transition temperatures are the guide for eye.
$T_{MI1}$ and $T_{MI2}$ reported in Ref.~\cite{Kotegawa2018} are also indicated by arrows.
}
\label{fig:fittingresults}
\end{figure}

Figure \ref{fig:fittingresults} shows the energy ($\hbar\omega_0$) of the lowest-energy Lorentz peak and the center of gravity of the spectra ($\langle\hbar\omega\rangle$) in $\hbar\omega \leq 1.2$~eV as a function of temperature.
In $E \parallel b_H$, both $\hbar\omega_0$ and $\langle\hbar\omega\rangle$ are almost constant above about $T_{MI2}$ ($\sim200$~K), 
but they increase with decreasing temperature below $T_{MI2}$, suggesting energy-gap opening.
On the other hand, in $E \parallel c_H$, the transition temperature is higher than that of $E \parallel b_H$, 
i.e., $\hbar\omega_0$ is almost constant above about $T_{MI1}$ ($\sim250$~K) but it increases below the temperature.
$\langle\hbar\omega\rangle$ has similar behavior but is almost constant above 230~K.
The energy gap in $E \parallel c_H$ is concluded to be formed below $240\pm10$~K.
Therefore, it is suggested that there is anisotropy in the temperature at which the energy gap begins to open.

The temperature dependences of $\hbar\omega_0$ and $\langle\hbar\omega\rangle$ can be fitted well by the gap function of the BCS theory, in which a pairing of charges is expected, and are plotted by thick dotted lines in Fig.~\ref{fig:fittingresults}.
This result is consistent with the previous study of CDW and spin-density wave (SDW) transitions such as that of (TMTSF)$_2$PF$_6$~\cite{Dressel1997} and CeOs$_2$Al$_{10}$~\cite{Kimura2011}.
Here, the gap opening temperatures is assumed as $205$~K in $E \parallel b_H$ and $240$~K in $E \parallel c_H$ (Vertical dashed lines in the figure).
As a result, both experimental curves of $\hbar\omega_0$ and $\langle\hbar\omega\rangle$ can be explained by the gap functions, i.e., charge pairing as well as the CDW formation works for the generation of the energy gap.

Finally, the origin of the anisotropy of the phase transition temperature is discussed.
According to the X-ray diffraction patterns in the LT and HT phases, the original unit cell in the HT phase, where all Ru atoms take uniform ionic state, 
forms the orthorhombic $3\times3\times3$ superlattice with nine sites with different valence in the LT phase~\cite{Kotegawa2018}.
Along the $c_H$ direction, every three sites in the HT phase becomes one unit cell in the LT phase with a long charge periodicity of six different Ru ions,
while, along the $b_H$ direction, a shorter charge periodicity of three different Ru ions appears, i.e.,
the different charge orderings along the $b_H$ and $c_H$ directions are expected.
According to the experimental results, the energy gap in $E \parallel c_H$ opens at higher temperature than that in $E \parallel b_H$.
This suggests that the ordering along the $c_H$ direction firstly occurs at the higher transition temperature of $T_{MI1}$, 
and subsequently the ordering along the $b_H$ direction occurs at the lower transition temperature of $T_{MI2}$.
Therefore, the two-step phase transition at $T_{MI1}$ and $T_{MI2}$ is considered to originate from the anisotropic ordering characters.
The ordering may not occur uniformly owing to the long charge ordering period, so incommensurate phase and domains may appear.

In RuP, two-step phase transition also appears, but the origin has not been revealed.
The crystal structure change through the two-step phase transition has not been clarified yet, in addition, anisotropic electronic structure as well as the anisotropic PE and $\sigma(\omega)$ spectra has not been reported.
To clarify the relation between the RuAs and RuP for the understanding of the similar materials, it is desirable to conduct detailed electronic structure studies.


\section{Conclusion}
To summarize, the change of the electronic structure due to the metal-to-insulator transition (MIT) of RuAs has been investigated by angle-integrated photoelectron (PE) and optical conductivity [$\sigma(\omega)$] spectra and LDA band calculations.
Both of PE and anisotropic $\sigma(\omega)$ spectra in the valence band region not only in the high-temperature (HT) phase above about $250$~K but also in the low-temperature (LT) phase below about $200$~K can be explained well by the LDA calculations.
Temperature-dependent PE intensity at $E_{\rm F}$ and Drude weight of $\sigma(\omega)$ spectra suggest that carrier density decreases from the HT phase to the LT phase owing to the MIT.
In $\sigma(\omega)$ spectra, anisotropic energy gap size in the LT phase and gap opening temperature have been observed and they are considered to be related to the anisotropic charge ordering.

\section*{Acknowledgments}
We would like to thank UVSOR staff members, especially F.~Teshima, K.~Tanaka and S.~Ideta, for their support during synchrotron radiation experiments.
Part of this work was performed under the Use-of-UVSOR Facility Program (Proposals Nos.~29-565, 29-829, 30-566, 30-576, 19-567) of the Institute for Molecular Science, National Institutes of Natural Sciences and Aichi Synchrotron Radiation Center, Aichi Science \& Technology Foundation.
This work was partly supported by JSPS KAKENHI (Grant No.~15H03676).


\end{document}